\documentclass{PoS}

\usepackage{amsmath,graphicx,verbatim,epsfig}

\usepackage[utf8x]{inputenc}
\usepackage{simplewick}

\newcommand{\nn}{\nonumber}

\newcommand{\bn}{{\overline n}}

\newcommand{\be}{\begin{equation}}
\newcommand{\ee}{\end{equation}}
\newcommand{\bea}{\begin{eqnarray}}
\newcommand{\eea}{\end{eqnarray}}
\newcommand{\balign}{\begin{align}}
\newcommand{\ealign}{\end{align}}
\newcommand{\as}{\alpha_s}

\newcommand{\bg}{\begin{gather}}
\newcommand{\foma}{\end{gather}}

\newcommand{\noopsort}[1]{}

\def\<{\langle}
\def\>{\rangle}

\def\a{\alpha}
\def\b{\beta}
\def\g{\gamma}  \def\G{\Gamma}

\def\m{\mu}

\def\({\left(}
\def\[{\left[}
\def\){\right)}
\def\]{\right]}

\def\ln{\hbox{ln}}

\def \le { \left    }
\def \ri { \right }

\title{Towards the Phenomenology of TMD Distributions at NNLL}
\ShortTitle{DIS2013}

\author{Miguel G. Echevarr\'ia\\
Departamento de F\'isica Te\'orica II,
Universidad Complutense de Madrid (UCM),
28040 Madrid, Spain\\
E-mail: \email{miguel.gechevarria@fis.ucm.es}}
\author{Ahmad Idilbi
\footnote{Current E-mail address: aui13@psu.edu}\\
European Centre for Theoretical Studies in Nuclear Physics and Related Areas (ECT*),
Villa Tambosi, Strada delle Tabarelle 286, I-38123, Villazzano, Trento, Italy
E-mail: \email{idilbi@ectstar.eu}}
\author{\speaker{Ignazio Scimemi}\\
Departamento de F\'isica Te\'orica II,
Universidad Complutense de Madrid (UCM),
28040 Madrid, Spain\\
E-mail: \email{ignazios@fis.ucm.es}}

\abstract{We discuss the recently proposed scheme for the evolution of Transverse Momentum Distributions  using the so-called ``$D^R$''.  
We discuss a new method to separate perturbative and non-perturbative effects in the analysis of the evolution of Transverse Momentum Distributions. 
We argue that the scheme is compatible with the standard CSS resummation approach when the running of the strong coupling is taken into account properly.}

\FullConference{
XXI International Worskhop on Deep-Inelastic Scattering and Related Subjects\\
April 22 - 26, 2013\\
Marseille, France}

\begin{document}

Modern high energy colliders are supposed to provide  a  great amount of information on observables which depend on transverse momentum of initial and/or final states. 
The analysis of this kind of observable relies on factorization theorems, which represent a unique tool to define non-perturbative observables and their evolution properties.
Recent works~\cite{GarciaEchevarria:2011rb,Echevarria:2012js,Collins:2011zzd} have achieved a definition of new non-perturbative  matrix elements,  the Transverse Momentum Dependent Parton Distribution Functions (TMDPDFs), which are obtained by considering factorization theorems either for the TMD Drell-Yan spectrum or the Semi-inclusive DIS process. 
The two approaches use two different settings for regularization (off-the-light-cone for~\cite{Collins:2011zzd} and on-the-light-cone, effective field theory and rapidity/infrared regulators for~\cite{GarciaEchevarria:2011rb,Echevarria:2012js}).

Despite many efforts, the practical implementation of evolution of TMDs is  still a source of discussion. 
As argued in~\cite{Echevarria:2012pw}, the anomalous dimension of the unpolarized TMDPDF can be given up to 3-loop order based on
a factorization theorem for $q_T$-dependent observables in a Drell-Yan process~\cite{GarciaEchevarria:2011rb,Echevarria:2012js}.
Such factorization theorem for the hadronic tensor can be written in impact parameter space, as
\begin{align}\label{eq:factth}
\tilde{M}(x_n,x_\bn,b;Q^2) &=
H(Q^2/\m^2)\,
\tilde{F}_n(x_n,b;\alpha Q^2,\m)\,
\tilde{F}_\bn(x_\bn,b;Q^2/\alpha,\m)
+ {\cal O}\le((bQ)^{-1}\ri)
\,.
\end{align}
where $H$ is the hard coefficient encoding the physics at the probing scale $Q$ and which is a polynomial of only $\ln(Q^2/\mu^2)$.
This quantity is built, to all orders in perturbation theory, by considering virtual Feynman diagrams only, and no real gluon emission has to be considered (even in diagrams with mixed real and gluon contributions). Moreover, the quantity $H$ has to be free from infrared physics, no matter how the latter is regularized. 
This is a general principle and it should work whether one works on- or off-the-light-cone.
The TMDs depend on the boost invariant real number  $\alpha$  defined in~\cite{Echevarria:2012js}  and  we choose $\alpha=1$.

Since the factorization theorem given above holds, at leading-twist, also for spin-dependent observables, one can apply the same arguments as for the unpolarized case, based on renormalization group invariance, to get a relation between the anomalous dimensions of the TMDs and the hard coefficient.
Then, the anomalous dimension for the TMD will be
\begin{align}
\g_F &=
-\frac{1}{2} \le[ 2\G_{\rm cusp}\, \ln\frac{Q^2}{\m^2} + 2\g^V \ri]
\,,
\end{align}
where $\g_H$ is known at 3-loop level~\cite{Idilbi:2006dg,Moch:2005id,Moch:2004pa} .
$\G_{\rm cusp}$ stands for the well-known cusp anomalous dimension in the fundamental representation.
This important result can be automatically extended to the eight leading-twist quark-TMDPDFs since the anomalous dimension is independent of spin structure.
As a result the evolution of a generic quark-TMDPDF is given by~\footnote{Since the evolution kernel is the same for $\tilde{F}_{n}$ and $\tilde{F}_{\bn}$, we have dropped out the $n,\bn$ labels.}
\begin{align}\label{eq:tmdevolution}
\tilde F(x,b;Q_f,\m_f)& = \tilde F(x,b;Q_i,\m_i)\, \tilde R(b;Q_i,\m_i,Q_f,\m_f)\,,
\end{align}
where the evolution kernel $\tilde R$ is~\cite{GarciaEchevarria:2011rb,Echevarria:2012js}
\begin{align}\label{eq:tmdkernel}
\tilde R(b;Q_i,\m_i,Q_f,\m_f) &=
\exp\le\{
\int_{\m_i}^{\m_f} \frac{d\bar\m}{\bar\m} \g_F\le(\as(\bar\m),\ln\frac{Q_f^2}{\bar\m^2} \ri)
\ri\}
\le( \frac{Q_f^2}{Q_i^2} \ri)^{-D\le(b;\m_i\ri)}
\,.
\end{align}
The $D$ term allows  the resummation of all $\ln(Q^2/q_T^2)$ by applying  the renormalization group invariance to the hadronic tensor $\tilde M$ in Eq.~(\ref{eq:factth}),
\begin{align}
\frac{dD}{d\ln\mu}=\G_{cusp}
\label{eq:drg}
\end{align}
and the $\G_{cusp}$ is known at 3-loops~\cite{Moch:2004pa}.
Matching the perturbative expansions of the $D$ term,
$D(b;\m) = \sum_{n=1}^\infty d_n(L_\perp) \left( \frac{\as}{4\pi} \right)^n\,,
\;
L_\perp=\ln\frac{\m^2 b^2}{4e^{-2\g_E}}
\,,
$ 
the cusp anomalous dimension $\G_{\rm cusp}$ and the QCD $\b$-function, one gets a recursive differential equation for $d_n$.
Solving this equation and  performing finally the all order resummation one gets 
\begin{align}\label{eq:resummedD}
 D^R &=
-\frac{\Gamma_0}{2\beta_0}\ln(1-X)
+ \frac{1}{2}\le(\frac{a}{1-X}\ri) \le[
- \frac{\beta_1\Gamma_0}{\beta_0^2} (X+\ln(1-X))
+\frac{\Gamma_1}{\beta_0} X\ri]
\nn\\
&+ \frac{1}{2}
\le(\frac{a}{1-X}\ri)^2\le[
2d_2(0)
+\frac{\Gamma_2}{2\beta_0}(X (2-X
))
+\frac{\beta_1\Gamma_1}{2 \beta_0^2} \le( X (X-2)-2 \ln (1-X)\ri)
+\frac{\beta_2\Gamma_0}{2\beta_0^2} X^2 
\ri.
\nn\\
&\le.
+\frac{\beta_1^2\Gamma_0}{2 \beta_0^3} (\ln^2(1-X)-X^2)
\ri]
 +...
\,,
\end{align}
where $a=\as/(4\pi)$, $X=a\b_0L_\perp$ and is valid for $X<1$.

The resummation using $D^R$ in  Eq.~(\ref{eq:resummedD}) and the usual CSS approach resum the same kind of logarithms, as can be straightforwardly shown before the insertion of any model or extra parameter in the CSS method.
In order to make the comparison easier, we also choose to refer to the case  $\mu_i=Q_i$ and $\mu_f=Q_f$  in Eq.~(\ref{eq:resummedD}).
In CSS approach the $D$ term is resummed using its RG-evolution in Eq.~(\ref{eq:drg}),
\begin{align}\label{eq:devolved0}
D\le(b;Q_i\ri) &=
D\le(b;\m_b\ri)
+ \int_{\m_b}^{Q_i}\frac{d\bar\m}{\bar\m} \G_{\rm cusp}
\,,
\end{align}
and  one chooses $\mu_b=2e^{-\gamma_E}/b$ to cancel the $L_\perp$ terms.
 At lowest order in perturbation theory  one gets
\begin{align}
\label{eq:CSSD1}
D(b; Q_i)&=-\frac{\Gamma_0}{ 2 \beta_0}\ln\frac{\alpha_s(Q_i)}{\alpha_s(\mu_b)}\ ,
\end{align}
and  finally re-expressing $\alpha_s(\mu_b)$ in terms  of $\alpha_s(Q_i)$,  at the proper perturbative order, $\alpha_s(\mu_b)=\alpha_s(Q_i)/(1-X)$, one finds
\begin{align}
\label{eq:CSSD2}
D(b;Q_i)&=-\frac{\Gamma_0}{ 2 \beta_0}\ln (1-X) \ ,
\end{align}
which coincides with the first term of the r.h.s of Eq.~(\ref{eq:resummedD}).
Repeating the same steps at higher orders one finally gets that the resummed $D$ in CSS approach, given in Eq.~(\ref{eq:devolved0}), and ours, given in Eq.~(\ref{eq:resummedD}), are exactly the same.
Thus, we conclude that $D^R$  coincides with the CSS evolution when all terms in CSS approach are resummed to their appropriate order.
However in practical implementation of CSS method one usually uses formulas like Eq.~(\ref{eq:CSSD1}), and so $\a_s(\mu_b)$ is affected  by the decoupling thresholds of c- and b-quarks, while in Eq.~(\ref{eq:CSSD2}) these thresholds marginally affect the final result.
The passage from Eq.~(\ref{eq:CSSD1}) to Eq.~(\ref{eq:CSSD2}) requires that no higher order contributions from the running of $\alpha_s$ are included and that the number of flavors included in the running of $\a_s(Q)$ and $\a_s(\mu_b)$ is the same.
The difference in the two formulas is clear if one inserts higher order contributions of the running of the coupling constant, together with the  implementation of the decoupling corrections at physical thresholds, as given in~\cite{Larin:1994va,Chetyrkin:1997un,Chetyrkin:2000yt,Chetyrkin:2005ia,Schroder:2005hy}.
In~\cite{Echevarria:2012pw} we checked that the solution provided by the $D^R$ is stable, while the direct use of Eq.~(\ref{eq:CSSD1}) leads to undesired divergent behavior for relatively low values of the impact parameter.
In other words, the implementation of $D^R$ takes into account the running of the coupling constant at the proper perturbative order and the decoupling of thresholds at the scale $Q_i$ automatically. 
In~\cite{Echevarria:2012pw} we discussed as well the decoupling of perturbative and non-perturbative effects in the analysis of TMDs. 
We suggest that a possible way to separate the two effects consists in writing the evolution kernel as 
\begin{align}\label{eq:Rcnp}
\tilde R(b;Q_i,\m_i,Q_f,\m_f) &=
\exp\le\{
\int_{\m_i}^{\m_f} \frac{d\bar\m}{\bar\m} \g_F\le(\as(\bar\m),\ln\frac{Q_f^2}{\bar\m^2} \ri)
\ri\}
\le( \frac{Q_f^2}{Q_i^2} \ri)^
{-\le[D^R\le(b;\m_i\ri)\theta(b_c-b)+D^{NP}(b;\m_i)\theta(b-b_c)\ri]}
\,,
\end{align}
where $b_c$ is some cutoff up to which $D^R$ converges and $D^{NP}$ a non-perturbative input for $b>b_c$. 
The advantage of this implementation of the kernel is that now the perturbative piece of the evolution is treated in a completely perturbative  way, independent from the non-perturbative input, which renders the results more predictive.
In~\cite{Echevarria:2012js} we discuss under which kinematical conditions the effect of $D^{NP}$ can be neglected, thus obtaining a parameter-free expression for the evolution kernel.

We conclude from this analysis that the use of $D^R$ and CSS  resummation differ in the treatment of thresholds and the running of the coupling constant, however both are consistent by construction.
A direct implementation of Eq.~(\ref{eq:devolved0}) with a running coupling  spoils the convergence of the resummation and results in an overestimate of higher order perturbative contributions.
The evident problems of Eq.~(\ref{eq:CSSD1}) are present also at NLL and NNLL. 
In the standard CSS approach one compensates for these higher order contributions by introducing some non-perturbative model. 
The correct perturbative expansion performed with our $D^R$ allows us to extend the region of the impact parameter space where it converges, or in other words, to obtain a parameter-free expression for the kernel in the perturbative domain, completely separating perturbative and non-perturbative effects.

\section*{Acknowdlegements}
This work is supported by the Spanish MEC, FPA2011-27853-CO2-02.
M.G.E. is supported by the PhD funding program of the Basque Country Government. The authors would like to thank Andreas Sch\"afer for his collaboration on the work presented in
\cite{Echevarria:2012pw}.



\begin{thebibliography}{99}


\bibitem{GarciaEchevarria:2011rb}
  M.~G.~Echevarria, A.~Idilbi and I.~Scimemi,
  JHEP {\bf 1207} (2012) 002
  [arXiv:1111.4996 [hep-ph]].

\bibitem{Echevarria:2012js}
  M.~G.~Echevarria, A.~Idilbi and I.~Scimemi,
  arXiv:1211.1947 [hep-ph].

\bibitem{Collins:2011zzd}
  J.~Collins,
  ``Foundations of perturbative QCD,''
  (Cambridge monographs on particle physics, nuclear physics and cosmology. 32)

\bibitem{Collins:2012uy}
  J.~C.~Collins and T.~C.~Rogers,
  arXiv:1210.2100 [hep-ph].

\bibitem{Echevarria:2012pw}
  M.~G.~Echevarria, A.~Idilbi, A.~Sch\"afer and I.~Scimemi,
  arXiv:1208.1281 [hep-ph].

\bibitem{Idilbi:2006dg}
  A.~Idilbi, X.~-d.~Ji and F.~Yuan,
  Nucl.\ Phys.\ B {\bf 753} (2006) 42
  [hep-ph/0605068].

\bibitem{Moch:2005id}
  S.~Moch, J.~A.~M.~Vermaseren and A.~Vogt,
  JHEP {\bf 0508} (2005) 049
  [hep-ph/0507039].

\bibitem{Moch:2004pa}
  S.~Moch, J.~A.~M.~Vermaseren and A.~Vogt,
  Nucl.\ Phys.\ B {\bf 688} (2004) 101
  [hep-ph/0403192].

\bibitem{Larin:1994va}
  S.~A.~Larin, T.~van Ritbergen and J.~A.~M.~Vermaseren,
  Nucl.\ Phys.\ B {\bf 438} (1995) 278
  [hep-ph/9411260].

\bibitem{Chetyrkin:1997un}
  K.~G.~Chetyrkin, B.~A.~Kniehl and M.~Steinhauser,
  Nucl.\ Phys.\ B {\bf 510} (1998) 61
  [hep-ph/9708255].

\bibitem{Chetyrkin:2000yt}
  K.~G.~Chetyrkin, J.~H.~Kuhn and M.~Steinhauser,
  Comput.\ Phys.\ Commun.\  {\bf 133} (2000) 43
  [hep-ph/0004189].

\bibitem{Chetyrkin:2005ia}
  K.~G.~Chetyrkin, J.~H.~Kuhn and C.~Sturm,
  Nucl.\ Phys.\ B {\bf 744} (2006) 121
  [hep-ph/0512060].

\bibitem{Schroder:2005hy}
  Y.~Schroder and M.~Steinhauser,
  JHEP {\bf 0601} (2006) 051
  [hep-ph/0512058].

\end{thebibliography}
\end{document}